\documentclass[aps,prl,two column]{revtex4-2}
\usepackage[utf8]{inputenc}
\usepackage[colorinlistoftodos, color=green!40, prependcaption]{todonotes}
\usepackage{amsthm}
\usepackage{mathtools}
\usepackage{physics}
\usepackage{xcolor}
\usepackage{graphicx}
\usepackage[left=23mm,right=13mm,top=35mm,columnsep=15pt]{geometry} 
\usepackage{adjustbox}
\usepackage{placeins}
\usepackage[T1]{fontenc}
\usepackage{lipsum}
\usepackage{csquotes}
\usepackage[pdftex, pdftitle={Article}, pdfauthor={Author}]{hyperref} 
\usepackage{braket}
\usepackage{siunitx}

\usepackage{xcolor} 
\usepackage{gensymb}
\DeclareSIUnit\bar{bar}
\DeclareSIUnit\cps{cps}

\begin{document}
\title{Frequency conversion to the telecom O-band using pressurized hydrogen}

\author{Anica Hamer}
    \affiliation{Physikalisches Institut, Rheinische Friedrich-Wilhelms-Universität Bonn, Bonn, Germany}
    
\author{Seyed Mahdi Razavi Tabar}
    \affiliation{Physikalisches Institut, Rheinische Friedrich-Wilhelms-Universität Bonn, Bonn, Germany}

\author{Priyanka Yashwantrao}
    \affiliation{Physikalisches Institut, Rheinische Friedrich-Wilhelms-Universität Bonn, Bonn, Germany}

\author{Alireza Aghababaei}
    \affiliation{Physikalisches Institut, Rheinische Friedrich-Wilhelms-Universität Bonn, Bonn, Germany}

\author{Frank Vewinger}
    \affiliation{Institut für Angewandte Physik, Rheinische Friedrich-Wilhelms-Universität Bonn, Bonn, Germany}

\author{Simon Stellmer}
    \email[Correspondence email address: ]{stellmer@uni-bonn.de}
    \affiliation{Physikalisches Institut, Rheinische Friedrich-Wilhelms-Universität Bonn, Bonn, Germany}

\date{December 23, 2023} 

\begin{abstract}
Large-scale quantum networks rely on optical fiber networks and photons as so-called flying qubits for information transport. While dispersion and absorption of optical fibers are minimum at the infrared telecom wavelengths, most atomic and solid state platforms operate at visible or near-infrared wavelengths. Quantum frequency conversion is required to bridge these two wavelength regimes, and nonlinear crystals are currently employed for this process. Here, we report on a novel approach of frequency conversion to the telecom band. This interaction is based on coherent Stokes Raman scattering (CSRS), a four-wave mixing process resonantly enhanced in a dense molecular hydrogen gas. We show the conversion of photons from \SI{863}{\nano\meter} to the telecom O-Band and demonstrate that the input polarization state is preserved. This process is intrinsically broad-band and can be adapted to any other wavelength.
\end{abstract}


\maketitle
\section{Introduction}

For quantum computing and communication purposes, state-preserving frequency conversion of single photons is indispensable in hybrid architecture with optical connections \cite{Silberhorn2017,Becher2012,Krutyanskiy2017,Fisher2021,Bell2017,Zhou2014,Tyumenev2022-sp}. The most widely used method of quantum frequency conversion is based on the $\chi^2$ nonlinearity of crystals \cite{Silberhorn2017, Becher2012, Fisher2021, Bell2017, Zhou2014}. This approach has been developed towards near-unit conversion efficiency and a drastically reduced incoherent background. In related work, the $\chi^3$ nonlinearity of atomic gases has also been employed in combination with short-pulse lasers \cite{Eramo1994,Tamaki1998,Babushkin2008}.

Still, this crystal-based approach comes with a range of limitations, including very narrow bandwidth, polarization-dependence of the conversion process, crystal absorption, and inevitable background emission.

In recent work, a new approach to frequency conversion based on resonantly enhanced nonlinear interaction in pressurized diatomic gases \cite{Tyumenev2022-sp,Aghababaei2023-pa} has been demonstrated, based on a coherent anti-Stokes Raman scattering (CARS) ``up-conversion" process in the UV spectral range. It was shown that this process is well suitable for single photon conversion in quantum information processing due to its intrinsically broad bandwidth and insensitivity to the input polarization. Furthermore, it is free of absorption losses, degradation, or power limitations. A wide selection of molecules, isotopic compositions, and transitions allows to cover a wide range of wavelengths.

Here, we present the extension of previous work to the corresponding "upconversion" process, known as coherent Stokes Raman scattering (CSRS) \cite{Zheng2015}, which also relies on a four-wave mixing process that is resonantly enhanced in the vicinity of a vibrational transition in a molecule.

In this letter, we demonstrate conversion from \SI{863}{\nano\meter} to \SI{1346}{\nano\meter}. These wavelengths are intentionally chosen to prepare for the conversion of the two entangled photons of the biexciton emission of indium arsenide (InAs)/gallium arsenide (GaAs) quantum dots \cite{Gao2012, Liu2019,Wang2019,Joecker2019} to the telecom O-band. This band includes the zero crossing of the dispersion in optical fibers and thus provides minimal signal distortion \cite{Winzer:18}. Connecting well-established solid-state platforms to optical fiber networks paves the way towards long-distance quantum communication \cite{Deutsch2023}.

\begin{figure*}[ht]
	\centering
	\includegraphics[width=\linewidth]{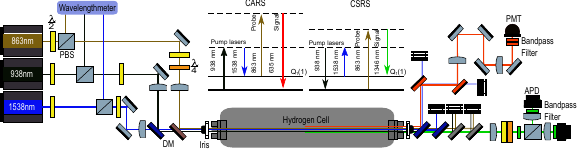}	
	\caption{Experimental setup, showing the pump fields at \SI{1538}{\nano\meter} and \SI{938}{\nano\meter} being overlapped with the signal input at \SI{863}{\nano\meter} via dichroic mirrors (DM) and focused into the hydrogen cell. After the cell the input light and the converted light are again separated via dichroic mirrors. The converted light of \SI{635}{\nano\meter} can be detected with one photomultiplier tube (PMT) in the visible range and two avalanche photon detectors (APDs) in the infrared. For the \SI{1346}{\nano\meter} the combined PBS and APDs setup allow for polarization-sensitive detection of the converted photons. The two different coherent Raman scattering processes are shown schematically.}
	\label{fig:setup}
\end{figure*}

\section{Experimental setup}

The concept of the experiment is shown in Fig. \ref{fig:setup}. The frequency conversion process is based on two pump fields whose frequency difference at matched to the vibrational level at \SI{124}{\tera\hertz} within the $Q_1(1)$ branch of molecular hydrogen. Here, the wavelengths of the two pump fields are chosen to be \SI{938}{\nano\meter} and \SI{1538}{\nano\meter}.

The \SI{938}{\nano\meter} laser is an amplified diode laser and provides \SI{0.65}{\watt} to the experiment. The other pump laser at \SI{1538}{\nano\meter} is a diode laser, boosted to \SI{15}{\watt} through a fiber amplifier. These two lasers are stabilized at the MHz-level by a wavelength meter.

The signal photons are derived from a diode laser operated near \SI{863}{\nano\meter} and attenuated to a typical power of a few \si{\micro\watt}. This wavelength is chosen to coincide with the emission of InAs/GaAs quantum dots. The beams are overlapped using dichroic mirrors, and focussed to waists of \SI{80}{\micro\meter} within the hydrogen cell. The beam powers are monitored, and the polarization of the beams can be set individually.

Our hydrogen cell is a home-built, cylindrical high-pressure vessel made from stainless steel that can stand a pressure of up to \SI{60}{\bar}. The cell has a \SI{12}{\milli\meter} bore and a length of \SI{140}{\milli\meter} and is capped by two uncoated sapphire windows.

Behind the cell, the converted photons at \SI{635}{\nano\meter} and \SI{1346}{\nano\meter} are separated using dichroic mirrors from the pump and signal light. Bandpass filters (\SI{10}{\nano\meter} width and \SI{93}{\percent} transmission) provide further spectral purification.

Photons converted to \SI{1346}{\nano\meter} pass through a polarizing beamsplitter (PBS) and are detected by two avalanche photodiodes (APD, ID Qube NIR, dead time \SI{100}{\nano\second}) with a quantum efficiency of \SI{12.5}{\percent}. Photons at \SI{635}{\nano\meter} follow a different path and are detected by a photomultiplier tube (PMT, Hamamatsu model H10682-210) with a quantum efficiency of \SI{4.5}{\percent}. Corresponding photon count rates are measured with a single-photon counting module with a dead time of \SI{6}{\nano\second}.

\section{Frequency Conversion}
\subsection{Resonance}
\begin{figure}[htpb]
	\centering
	\includegraphics[width=\linewidth]{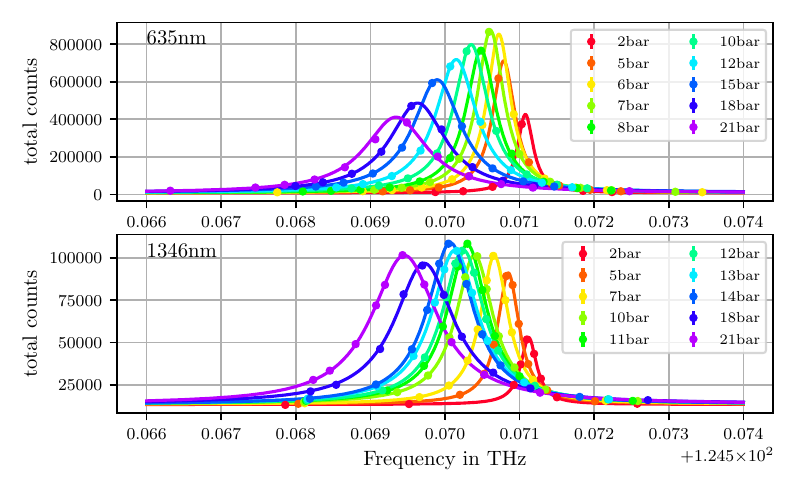}
	\caption{Pressure-dependence of the CSRS process (conversion to \SI{635}{\nano\meter}, top) and of the CARS process (to \SI{1346}{\nano\meter}, bottom) near the $Q_1(1)$ transition around \SI{124}{\tera\hertz}.}
	\label{fig:pressure}
\end{figure}

\begin{figure}[htpb]
	\centering
	\includegraphics[width=\linewidth]{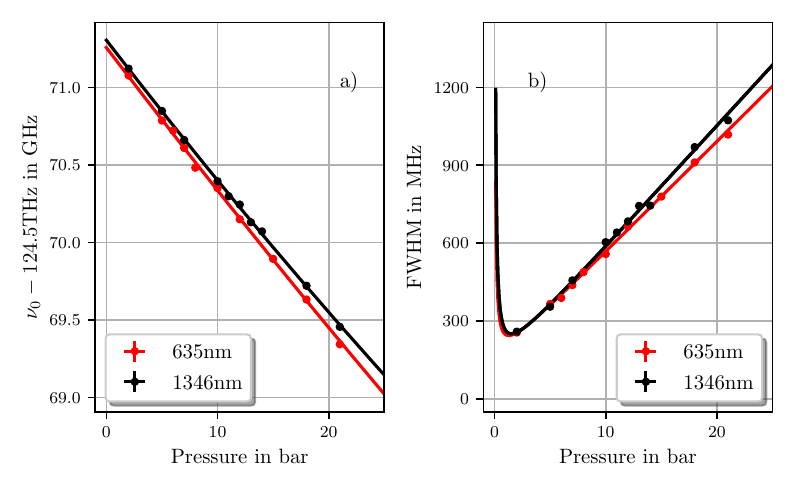}
	\caption{Interpolation of the extracted center positions of the Lorentzian curves depending on the pressure and separated for the two frequency-converted wavelengths.}
	\label{fig:position}
\end{figure}

We scan the relative detuning of the pump fields around the molecular resonance and observe frequency conversion simultaneously through the CRSR process at \SI{1346}{\nano\meter} and through the CARS process at \SI{635}{\nano\meter}. Resonance lineshapes and efficiencies depend strongly on the hydrogen pressure; see Fig.~\ref{fig:pressure}. By fitting resonances with Lorentzian curves we obtain the pressure-dependent center positions and widths are given in Fig.~\ref{fig:position}.

From our data, we derive a pressure (collisonal) shift of d$\nu$/dP = \SI{-94 \pm 1}{\mega\hertz/\bar} for the CARS process and \SI{-93 \pm 1}{\mega\hertz/\bar} for the CSRS process, which are in the interval of all measured data sets reported earlier \cite{Bischel86,Looi1978}. The uncertainty is entirely dominated in the uncertainty of the pressure gauge. Extrapolating the curves to zero pressure gives an unperturbed resonance frequency of $\nu(0)$ = \SI{124.571257 \pm 0.000002}{\tera\hertz} (CARS) and $\nu(0)$ = \SI{124.571304 \pm 0.000002}{\tera\hertz} (CSRS), in agreement with earlier data \cite{Looi1978}. Here, the uncertainty is dominated by the accuracy of the wavelength meter.

Even for the smallest pressures evaluated here, the linewidth of the unperturbed molecular resonance is narrowed dramatically through the Dicke effect \cite{Bischel86}. For pressures above about 1 bar, the linewidth increases approximately linearly through density (pressure) broadening. The data shown in Fig.~\ref{fig:position} is approximated by the model described in Ref.~\cite{Bischel86} and yields a pressure-dependent linewidth of \SI{42.7 \pm 0.5}{\mega\hertz\per\bar} for the CARS process and \SI{46.9 \pm 0.5}{\mega\hertz\per\bar} for the CSRS process, compatible with earlier studies \cite{Bischel86,Welsh1961}. 

\subsection{Efficiency}
\begin{figure}[hptb]
	\centering
	\includegraphics[width=\linewidth]{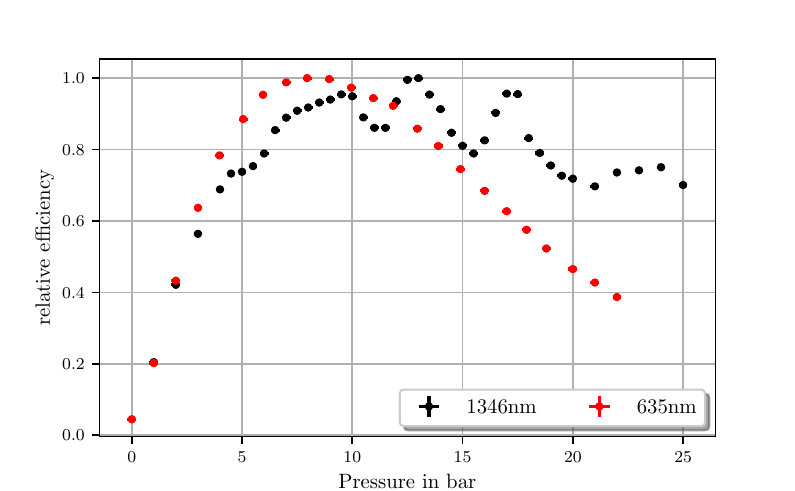}
	\caption{Conversion efficiency depending on the pressure of the hydrogen gas based on the the phase mismatch between the light fields. The data is normalized to its maximum value.}
	\label{fig:eff}
\end{figure}

As shown already in Fig.~\ref{fig:pressure}, the conversion efficiency is pressure-dependent. This dependence is shown in Fig.~\ref{fig:eff}. In the CARS process to \SI{635}{\nano\meter}, a clear maximum near \SI{8 \pm 0.2}{\bar} can be observed. For our experimental parameters stated above, the internal conversion efficiency is \SI{8.1e-10}{} and reduces to an external efficiency of \SI{1.5 e-11}{}, which takes into account the transmission losses of the optics and the \SI{4.5}{\percent} efficiency of the detector.

The behaviour of the CSRS process to \SI{1346}{\nano\meter} is strikingly different. The efficiency does not show a pronounced maximum, but oscillates around an internal efficiency of about \SI{1.1e-9}{} for a pressure between about 5 and \SI{20}{\bar} (external efficiency: \SI{9.0e-11}{}). We attribute the oscillations to fluctuations in the experimental parameters during the measurements, a behaviour we have already observed earlier, as each data point corresponds to a measurement time of \SI{2}{\minute} \cite{Aghababaei2023-pa}.

The noise level, \textit{i.e.,} the rate of undesired background photons originating from the interaction of the pump fields, is compatible with zero within our experimental uncertainty.

\subsection{Preservation of polarization}

Quantum frequency conversion requires a faithful preservation of the initial photon state, namely its polarization. In contrast to a crystal, the hydrogen gas itself is isotropic, and the reference frame is set by the light field. To test the preservation of polarization, the output state is read out for two polarization bases, namely linear and circular.

As the preservation of polarization in CARS processes has already been shown \cite{Aghababaei2023-pa}, we only investigate the CSRS process here. The detection unit involves waveplates at \SI{1346}{\nano\meter}, a polarization beamsplitter (PBS), and detectors on the two outputs.

\subsubsection{Linear polarization}

\begin{figure}
	\centering
	\includegraphics[width=\linewidth]{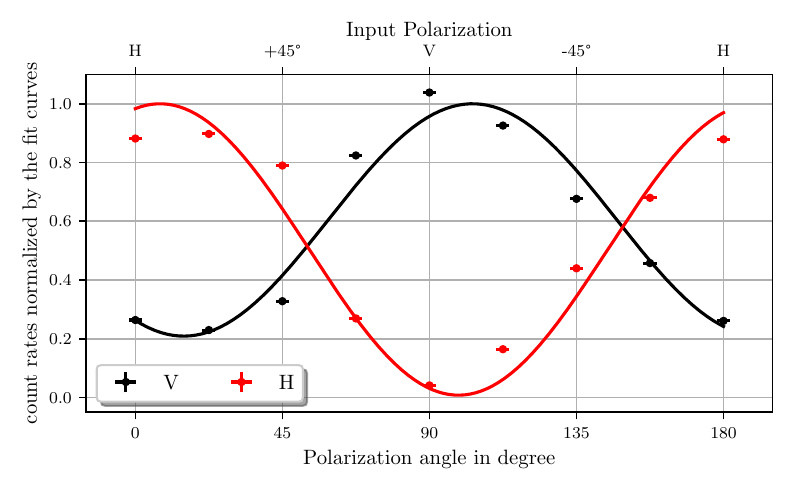}
	\caption{Normalized measurement of the Stokes parameter set of linear and diagonal with the highest values per each detector. Fitted via two sine functions. Error bars for the y-axis are determined by the square root of the count rate and are too small to be noticeable in this depiction. }
	\label{fig:HD}
\end{figure}

To check for the preservation of the linear polarization, the pump polarization is set to horizontal and the polarization of the signal photons is rotated. The recorded polarization state of the converted photons shown in Fig.~\ref{fig:HD} faithfully follows the input polarization.

Deviations can be attributed to imperfect waveplates and beamsplitters, as well as leakage light from the pump fields.

\subsubsection{Circular polarization}
To check for preservation of the phase, the polarization of the signal photons is turned from linear to circular. Correspondingly, the circularly polarized signal photons at \SI{1346}{\nano\meter} polarized diagonally before the detectors.

\begin{figure}
	\centering
	\includegraphics[width=\linewidth]{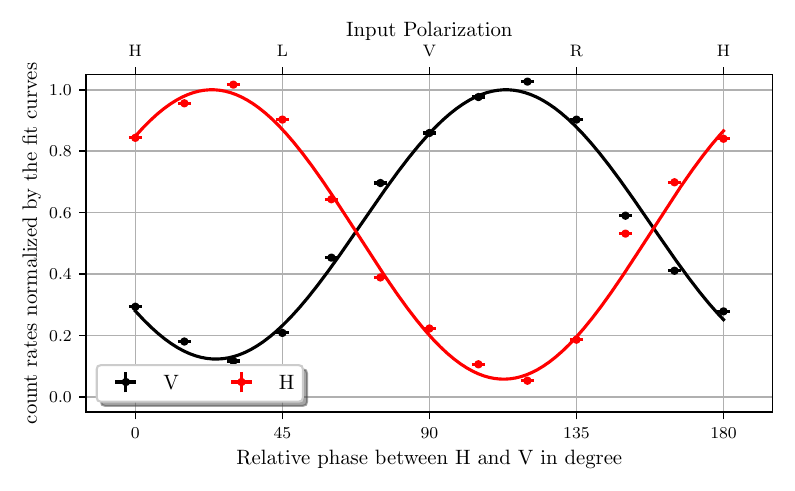}
	\caption{Normalized measurement of the Stokes parameter set of linear and circular with the highest values per each detector. Fitted via two sine functions. Error bars for the y-axis are determined by the square root of the count rate and are too small to be noticeable in this depiction.}
	\label{fig:HC}
\end{figure}

The results are shown in Fig.~\ref{fig:HC}. Again the output polarization faithfully follows the input polarization, and the small deviations from an ideal behaviour can be fully attributed to imperfect polarization optics.
  
\subsubsection{Fidelity}
We fit sine functions to the data sets presented in Figs.~\ref{fig:HD} and ~\ref{fig:HC} and compute the contrast for each measurement. The contrast for these four data sets ranges between 79.1\% and 99.3\% and is limited by the background signal and imperfect polarization optics, which are different for the two detectors and which we do not correct for. Here, we determine the fidelity as the average of the four contrast values and arrive at a fidelity of 0.904.

\section{Conclusion} 

In this work, we have further developed our novel approach to frequency conversion. While our previous work focused on a CARS process in the UV range \cite{Aghababaei2023-pa}, we here presented a study on a CSRS process in the NIR range.

We have demonstrated frequency conversion from 863\,nm to a telecom wavelength and have shown the a photon's polarization state is preserved. The poor efficiency of this preliminary study can be readily increased though tighter focusing of the beams, increase of the pump powers, and better mode overlap. We anticipate an efficiency increase by orders of magnitude by employing hydrogen-filled hollow-core fibers. Near-unit efficiency has recently been shown in closely related work at comparable wavelengths \cite{Tyumenev2022-sp}. Dedicated polarization optics and improved spectral filters will reduce the background and thereby increase the process fidelity.

\section{backmatter}
\subsection{Funding}
We acknowledge funding by Deutsche Forschungsgemeinschaft DFG through grant INST 217/978-1 FUGG and through the Cluster of Excellence ML4Q (EXC 2004/1 – 390534769), as well as funding by BMBF through the QuantERA project QuantumGuide.

\subsection{Acknowledgments}
We thank all members of the Cluster of Excellence ML4Q, as well as Michael Frosz from the MPI for the Science of Light, for stimulating discussions.


\subsection{Data Availability Statement}
Data underlying the results presented in this paper are not publicly available at this time but may be obtained from the authors upon reasonable request.


\bibliography{bib}

\end{document}